\documentclass[aps,twocolumn,groupedaddress,showpacs,floats,amssymb]{revtex4-1}
\usepackage[dvips]{graphicx}
\usepackage{bm}

\def\he4{$^4$He}
\def\hel3{$^3$He}
\def\Am3{\AA$^{-3}$}
\def\beq{\begin{equation}}
\def\eeq{\end{equation}}

\begin{document}

\noindent {\bf Comment on ``Dislocation Structure and Mobility in hcp $^4$He"}\vspace{2mm}\\



In their Letter~\cite{Borda},  Borda, Cai, and de Koning report the results of {\it ab initio} simulations of dislocations responsible for the giant plasticity \cite{Balibar}.  
The authors claim key insights into  the recent interpretations of (i) the giant plasticity  and (ii) the mass flow junction experiments.  
The purpose of this Comment is clarifying the role of dislocations in the mass flow in conjunction with explaining that the part (ii) of the claim is misleading.

Borda {\it et al.} find that their dislocations  do not have superfluid cores.  This fact, however, is {\it not} crucial for the interpretation of the mass supertransport,  
including the effect of giant isochoric compressibility (aka the syringe effect) \cite{Hallock2009}.  Furthermore, the fact is not even new. 

The first-principle theoretical results showing that certain (!) screw and edge dislocations feature superfluid cores were reported in Refs.~\cite{screw} and \cite{sclimb}, respectively. Also, the behavior of the edge dislocation of Ref.~\cite{sclimb} was found to be consistent with the phenomenon of superclimb. In its turn, the superclimb remains the only known underlying mechanism behind the syringe effect. Apart from being fundamentally interesting on its own, the syringe effect is central for the liquidless  supertransport setups \cite{Beamish2016}. Thus, any theory aiming at explaining the supertransport through solid must account for the syringe effect too. The results of Ref.~\cite{screw,sclimb} provide a consistent, and up to now unique, first-principle basis for interpreting all known supertransport-related  phenomena in solid \he4  \cite{Hallock2009,Hallock2012,Hallock2014,Beamish2016}.

As is known, dislocations are characterized by orientation of their core and the Burgers vector. The dislocations studied in Ref.~\cite{Borda}---with the core and Burgers vector both {\it along} basal plane---are different from those found to have superfluid core \cite{screw,sclimb}---with the Burgers vector {\it perpendicular} to the basal plane and core either perpendicular to the plane \cite{screw} or along the plane \cite{sclimb}. Hence, the statement that (Qt) ``the interpretation of recent mass flow experiments in terms of a network of 1D Luttinger-liquid systems in the form of superfluid dislocation cores does not involve basal-plane dislocations" made in Ref.~\cite{Borda} is misleading.  

Moreover, that cores of dislocations with the Burgers vector along basal plane (studied in Ref.~\cite{Borda}) are not superfluid has been emphasized in Refs.~\cite{GB,screw,stress}: in the caption to Fig.~6 in Ref.~\cite{GB}; on p.1 at the end of the 1st and  beginning of the 2nd column and on p.3 of Ref.~\cite{screw}, (Qt) ``[In the case of edge dislocations, this protocol leads to an insulating ground state.]"; in the last paragraph on p.3 of Ref.~\cite{stress}. Likewise, the effect of splitting into partials---claimed in Ref.~\cite{Borda} as a new and crucial observation---has been reported in Refs.~\cite{stress,sclimb} for the edge dislocations of both types---with  (in the section ``Numerical results" on p.3 of Ref.~\cite{sclimb})  and without superfluid cores  (in the the last paragraph on p.3 of Ref.~\cite{stress}).   Furthermore, the energy of the structural fault found in Ref.~\cite{sclimb} to be much smaller than any other typical energy scale in solid \he4 implies large splitting of dislocations with core in the basal plane. Thus,  Ref.~\cite{Borda} neither negates the results of Refs.~\cite{GB,screw,stress,sclimb} nor provides new insights into superfluidity of dislocations.

Finally, it is worth noting that the observation of no superfluidity in Ref.~\cite{Borda} lacks not only novelty but, most likely, also the control of the numerical data. While dealing with large systems, the authors
do not use worm updates \cite{WA}. In such a case, PIMC algorithm  is  known to be notoriously prone to non-ergodicity in the worldline winding number space, so that the absence of macroscopic 
permutation cycles could merely reflect the non-ergodicity of the scheme rather than the absence of superfluidity.

We thank Nikolay Prokof'ev for useful discussions and acknowledge support from the National Science Foundation under the grants PHY-1314469 and PHY-1314735.
\vspace{2mm}\\
{\small \noindent   A. B. Kuklov$^1$ and B. V. Svistunov$^2$\\
$^1$Department of Engineering Science and Physics,
CUNY, Staten Island, NY 10314.\\
$^2$Department of Physics,  University of Massachusetts, Amherst, MA 01003.}
\bigskip

\end{document}